\documentstyle[12pt,epsf]{article}
\begin{document}
\begin{center}
{\large\bf Chiral Phase Transitions in QCD at Finite Temperature: \\
Hard-Thermal-Loop Resummed Dyson-Schwinger Equation \\
in the Real Time Formalism} 
\end{center}

\vspace{0.3cm}
\begin{center}
Hisao NAKKAGAWA, Hiroshi YOKOTA, Koji YOSHIDA \\
{\it Institute for Natural Sciences, Nara University, Nara 631-8502, Japan}
\end{center}
\begin{center}
and Yuko FUEKI \\
{\it Department of Physics,  Nara Women's University,
   Nara 630-8506, Japan}
\end{center}

\vspace{0.5cm}
\begin{center}
\begin{minipage}{13cm}
{\small
Chiral phase transition in thermal QCD is studied by using
the Dyson-Schwinger (DS) equation in the real time hard thermal loop
approximation. Our results on the critical temperature and the
critical coupling are significantly different from those in the
preceding analyses in the ladder DS equation, showing the
importance of properly taking into account the essential thermal
effects, namely the Landau damping and the unstable nature of thermal
quasiparticles.}
\end{minipage}
\end{center}

\vspace{1cm}
Although lots of efforts 
have been made to undestand the 
temperature/density-dependent phase transition in thermal QCD, 
we still have not really understood yet even the relation between
the chiral and the deconfinement phase transitions. 
The analysis with the use of Dyson-Schwinger (DS) equation is a
powerful tool to investigate {\it with the analytic procedure} 
the phase structure of gauge theories. At finite temperature/density,
however, we should seriously ask whether the preceding results [1] 
of the DS equation analysis on the chiral phase transition could be
relied on being the real consequences of thermal gauge 
theories. This is because in previous analyses the lessons from vacuum 
theories have been applied without close examination, 
having missed the essential thermal effects, thus causing the neglection
of large contributions to the DS equation unless 
otherwise existed. 

We here make a re-analysis by studying the hard-thermal-loop
(HTL) resummed DS equation in the real time formalism,
giving a new understanding on the phase structure and the mechanism 
of phase transition in thermal gauge theories. Main interest of the
present investigation lies in clarifying what are the essential
temperature effects that govern the phase transition and also in
finding how we can closely take these  effects into the "kernel" 
of the DS equation. Essential procedures of our analysis can be 
summerised as follows;

i) We use the real time closed-time-path formalism and study the
physical mass function $\Sigma_R$ itself, not the $\Sigma_{11}$, 
of the retarded fermion propagator. $\Sigma_R$ is the mass 
function of "unstable" quasi-particle in thermal field theories, 
thus having non-trivial imaginary parts as well as non-trivial 
wave-function renormalization constants.

ii) To closely estimating the dominant temperature effects we focus
on studying the DS equation being exact up to HTL approximation: 
both the gauge boson propagator and the vertex functions are 
determined within the HTL approximation, guaranteeing the gauge
invariance of the result at least in the perturbative analysis. 
With the HTL vertex functions [2] we can derive the desired HTL
resummed DS equation.

The second point listed above is better to be taken step by step 
into the actual analysis of the DS equation. In the present paper 
we present the result of our first step investigation; focussing 
on what happens when taking into account exactly the HTL resummed 
gauge boson propagators. We show explicitly the results of 
preceding analyses can not be the real consequences of thermal 
gauge field theories. 

Decomposing the fermion mass function $\Sigma_R$ as
\begin{equation}
 \Sigma_R(P)=(1-A(P))p_i\gamma^i -B(P)\gamma^0+C(P) \ ,
\end{equation}
with  $A(P)$, $B(P)$ and $C(P)$ being the three independent scalar 
invariants to be determined, the DS equation we here investigate 
in the point vertex approximation with the exact HTL resummed gauge
boson propagator becomes as ($P^{\mu}=(p_0, {\bf p})$), 
\begin{eqnarray}
  & & -p^2[1-A(P)] = -g^2 C_F \left. \int \frac{d^4K}{(2 \pi)^4}
       \right[ \{1+2n_B(p_0-k_0) \} Im[\ ^*G^{\rho \sigma}_R(P-K)]
       \times  \nonumber \\
  & & \Bigl[ \{ K_{\sigma}P_{\rho} + K_{\rho} P_{\sigma}
       - p_0 (K_{\sigma} g_{\rho 0} + K_{\rho} g_{\sigma 0} ) 
       - k_0 (P_{\sigma} g_{\rho 0} + P_{\rho} g_{\sigma 0} )
       + pkz g_{\sigma \rho} \nonumber \\
  & & + 2p_0k_0g_{\sigma 0}g_{\rho 0} \}\frac{A(K)}{[k_0+B(K)+i
       \epsilon]^2 - A(K)^2k^2 -C(K)^2 }
       + \{ P_{\sigma} g_{\rho 0} + P_{\rho} g_{\sigma 0} \nonumber \\
  & & - 2p_0 g_{\sigma 0} g_{\rho 0} \}
       \frac{k_0+B(K)}{[k_0+B(K)+i \epsilon]^2 - A(K)^2k^2
       -C(K)^2 } \Bigr] + \{1-2n_F(k_0) \}
       \times \nonumber \\ 
  & & \ ^*G^{\rho \sigma}_R(P-K) Im \Bigl[
       \{ K_{\sigma}P_{\rho}  + K_{\rho} P_{\sigma} - p_0 (K_{\sigma}
       g_{\rho 0} + K_{\rho} g_{\sigma 0} ) - k_0 (P_{\sigma}
       g_{\rho 0} + P_{\rho} g_{\sigma 0} ) \nonumber \\
  & & + pkz g_{\sigma \rho} + 2p_0k_0g_{\sigma 0}g_{\rho 0}\}
       \frac{A(K)}{[k_0+B(K)+i \epsilon]^2 - A(K)^2k^2-C(K)^2 } 
       \nonumber \\
  & & \left. +  \{ P_{\sigma} g_{\rho 0} + P_{\rho} g_{\sigma 0}
       - 2p_0 g_{\sigma 0} g_{\rho 0} \}
       \frac{k_0+B(K)}{[k_0+B(K)+i \epsilon]^2 - A(K)^2k^2
       -C(K)^2 } \Bigr] \right] \ , \\
  & &  - B(P)= -g^2 C_F \left. \int \frac{d^4K}{(2 \pi)^4} \right[
        \{1+2_B(p_0-k_0)\} Im[\ ^*G^{\rho \sigma}_R(P-K)] \times
         \nonumber \\
  & & \Bigl[ \{ K_{\sigma} g_{\rho 0} + K_{\rho} g_{\sigma 0}
       - 2k_0 g_{\sigma 0} g_{\rho 0} \}
       \frac{A(K)}{[k_0+B(K)+i \epsilon]^2 - A(K)^2k^2
       -C(K)^2 } \nonumber \\
  & & + \{ 2g_{\rho 0} 2g_{\sigma 0} - g_{\sigma \rho} \} 
       \frac{k_0+B(K)}{[k_0+B(K)+i \epsilon]^2 - A(K)^2k^2-C(K)^2 }
       \Bigr] + \{1-2n_F(k_0) \} \times \nonumber \\ 
  & & \ ^*G^{\rho \sigma}_R(P-K) Im \Bigl[ \frac{A(K)}{[k_0+B(K)+i
       \epsilon]^2 - A(K)^2k^2 -C(K)^2 } 
       \{ K_{\sigma} g_{\rho 0} + K_{\rho} g_{\sigma 0}  \nonumber \\
  & & \left. - 2k_0 g_{\sigma 0} g_{\rho 0} \} + \frac{k_0+B(K)}{[k_0+B(K)+
       i \epsilon]^2 - A(K)^2k^2-C(K)^2 }
       \{ 2g_{\rho 0} 2g_{\sigma 0} - g_{\sigma \rho} \} \Bigr] 
       \right] \ , \\
  & & C(P)= -g^2 C_F \int \frac{d^4K}{(2 \pi)^4} g_{\sigma \rho} 
       \{1+2_B(p_0-k_0) \} Im[\ ^*G^{\rho \sigma}_R(P-K)]
       \times \nonumber \\
  & & \Bigl[ \frac{C(K)}{[k_0+B(K)+i \epsilon]^2 - A(K)^2k^2
       -C(K)^2 } + \{1-2n_F(k_0) \} \times \nonumber \\ 
  & & \left. \ ^*G^{\rho \sigma}_R(P-K) Im \Bigl[
       \frac{C(K)}{[k_0+B(K)+i \epsilon]^2 - A(K)^2k^2
       -C(K)^2 } \Bigr] \right] \ ,
\end{eqnarray}
where the cutoff scale $\Lambda$ is introduced, regularizing the integral.
$\ ^{\ast}G^{\rho \sigma}_R(K)$ in Eqs.(2-4) is the HTL
resummed retarded gauge boson propagator [3],
\begin{equation}
\! \! \! \! \! \! \! \! \! \! \! \! \! \! \! \! \! \! \! \! \! \! \!
\! \! \! \!
  \ ^*G^{\rho \sigma}_R(K)= \frac{1}{\ ^{\ast}\Pi^R_T-K^2-i \epsilon k_0}
  A^{\rho \sigma} + \frac{1}{\ ^{\ast}\Pi^R_L-K^2-i \epsilon k_0}
  B^{\rho \sigma} - \frac{\xi}{K^2+i \epsilon k_0}
  D^{\rho \sigma},
\end{equation}
where $\ ^{\ast}\Pi^R_{T/L}(K)$ being the HTL resummed 
transverse/longitudinal mass function [4], and $A^{\rho \sigma}=g^{\rho \sigma}
- B^{\rho \sigma}- D^{\rho \sigma}$,
$B^{\rho \sigma}=- \tilde{K}^{\rho} \tilde{K}^{\sigma}/K^2$, 
$D^{\rho \sigma}= K^{\rho} K^{\sigma}/K^2$, being the projection tensors.
In the above, $\tilde{K}=(k, k_0{\bf \hat{k}})$, $k=\sqrt{{\bf k}^2}$
and ${\bf \hat{k}}={\bf k}/k$.

The DS equation, Eqs.(2-4), is still quite tough to be attacked,
forcing us further approximations for the analysis to be effectively
carried out. However, the approximation made use of must be 
consistent with the HTL approximation, without missing the 
important thermal effects out of the kernel of the DS equation.

Here it is worth noticing that the instantaneous exchange (IE) 
approximation frequently used in the preceding analyses [1] is
{\it not} compatible with the HTL approximation in the strict sense.
In the exact IE-limit the HTL resummed transverse mass function, 
$\ ^{\ast}\Pi^R_T(K)$, vanishes and the transverse
(magnetic) mode becomes totally massless. Namely the IE 
approximation discards the important thermal effect coming from
the Landau damping, thus dismissing the dynamical screening of 
the magnetic mode, causing the famous quadratic divergence of the
Rutherford scattering cross section. This can be clearly seen by 
taking the IE-limit of the DS equation, Eqs.(2-4), and neglecting
$Im[A(P)]$ and $Im[C(P)]$, then obtaining the following equation 
for $Im[B(P)]$ ($E \equiv \sqrt{ ({\bf p}-{\bf k})^2 } $),
\begin{equation}
\! \! \! \! \!
Im[B(P)] = \frac{g^2C_F}{4 \pi} m_g^2 T \int_0^{\infty} k^2 dk \int_{-1}^1
 dz \left( \frac{1}{E[E^2+m_g^2]^2} + \frac{1}{E^5} \right) ,
\end{equation}
showing $Im[B(P)]$ to be quadratically divergent. The reason why in
the previous analyses this divergence problem did not appear is 
that the imaginary part of $\Sigma_R$ is completely neglected 
there from the beginning, namely the equation for $Im[\Sigma_R]$ 
is totally discarded. 

Taking the above into account the approximation we further make use
of is the IE approximation to the longitudinal gauge boson 
propagator, by keeping the exact HTL resummed transverse 
propagator. In the IE-limit the HTL resummed longitudinal mass
function, $\ ^{\ast}\Pi^R_L(K)$, has a nonvanishing thermal 
Debye screening mass $m_g^2
\sim (gT)^2$, thus even in the IE limit the longitudinal mode can 
take into account the essential thermal effect. In the present 
analysis the gauge is fixed to the Landau gauge ($\xi =0$).

Now we solve the DS equations Eqs.(2-4) with 
the IE approximation to the longitudinal mode, using the iterating
procedure with suitable trial functions for the
solution. This method 
is useful so long as the convergence of the iteration is guaranteed. 

At each iteration, the three-fold integration is performed over $k$,
$k_0$ and $z=\cos\theta$, $\theta=\hat{\bf p} \cdot \hat{\bf
k}$. The integration kernel of the
present DS equation shows a little bit singular behavior, and the
numerical integration of such a singular integrand needs careful 
integration prescription, which is properly managed. As a
result we performed the 1000 times iterations, obtaining fairly 
stable solutions. 

Result of the present analysis shows that the wave function 
renormalization constants receive 10-20 percent corrections, 
indicating the necessity of gauge-parameter dependent analysis. 
The generated size of the imaginary part is nearly the same as 
that of the real part, indicating the existence of the non-trivial
imaginary parts, which is as expected though completely neglected
in the previous analyses. 

Let us now show the behavior of the mass function at some fixed 
value of $p$ as a function of the parameters $\alpha=g^2C_F/4\pi$
and $T$. The $T$ dependence of the mass function $Re[C(P)]$
for various fixed values of 
$\alpha$, and 
the $\alpha$ dependence of $Re[C(P)]$ for various fixed
values of $T$ are shown in Figs.1a and 1b, both with $p_0=0$ and
$p=0.1\Lambda$. The errors resulting from the fluctuations are 
smaller than the size of the symbol used for each sample point in
Figs.1a and 1b. From these figures we can see the two facts; 
i) The chiral phase transition is of second order,  since a 
fermion mass is generated at a critical value of the 
temperature $T$ or at the critical coupling constant $\alpha$ 
without any discontinuity, and ii) the critical temperature 
$T_c$ at fixed value of $\alpha$ is significantly higher than
the previous results [1], namely the restoration of chiral 
symmetry occurs at higher temperature than previously expected. 
Our result shows that as $T$ becomes smaller the critical coupling
constant $\alpha_c$ also becomes smaller, being consistent with the
zero temperature result, where the critical coupling constant 
$\alpha^0_c = \pi/3$.
The second fact clearly shows that in the previous analyses the
important temperature effects are neglected due to the inappropriate
approximations.

Present result shows the correctness of our research-strategy, 
thus the importance of the full HTL resummed DS equation analysis 
of the chiral phase transition at finite temperature/density, 
which is now under investigation.  

\begin{figure}[htbp]
\begin{center}
\epsfxsize=8cm
\epsfbox{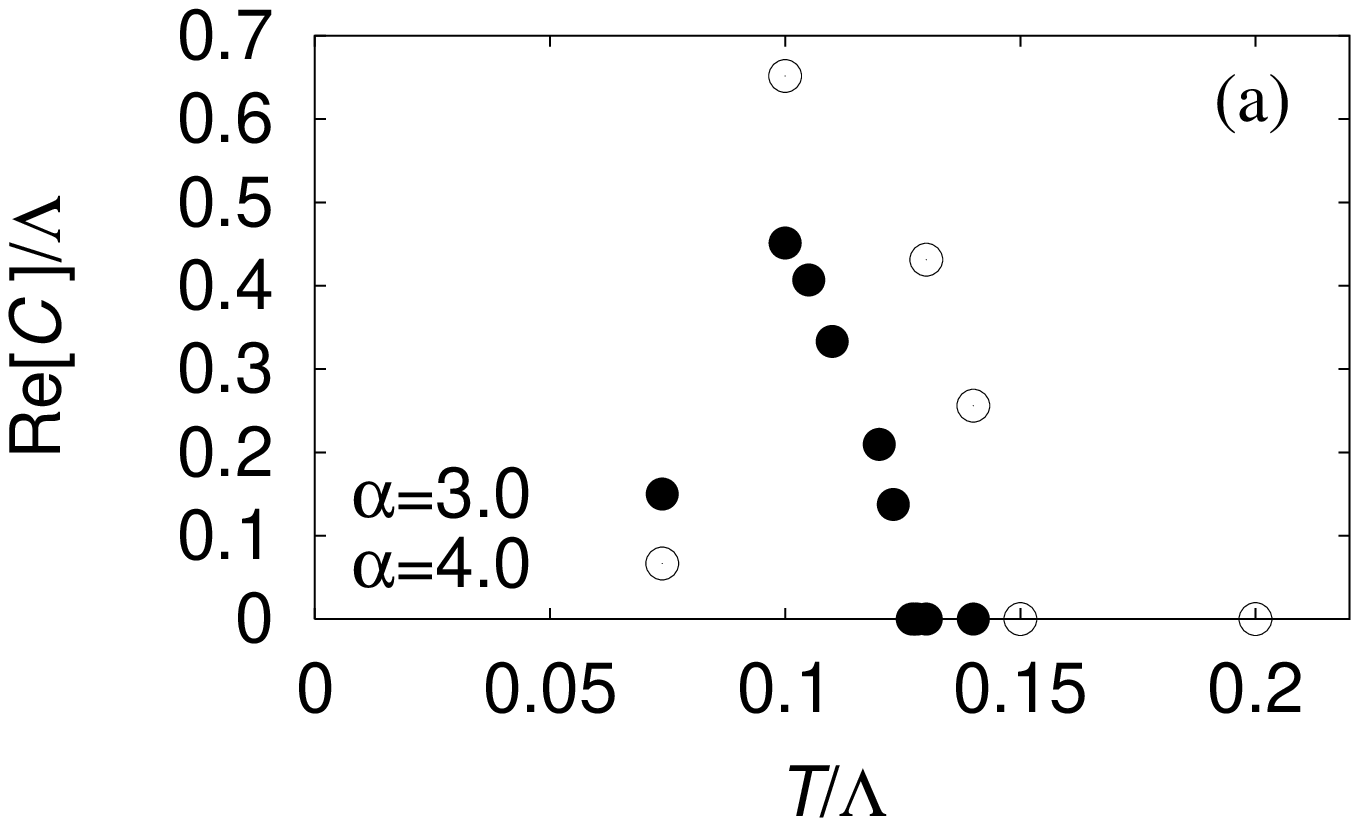}
\epsfxsize=8cm
\epsfbox{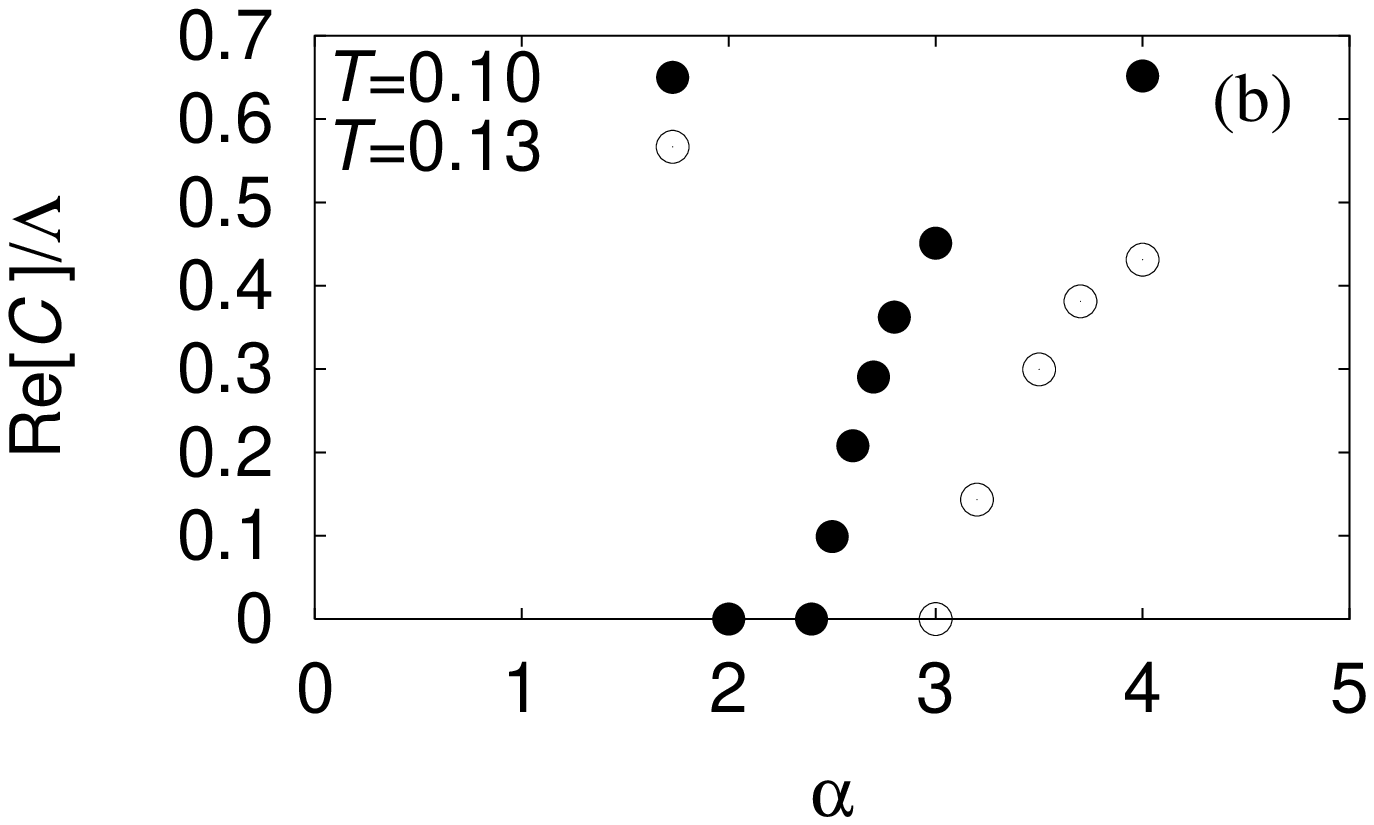}
\end{center}
\caption{}
\end{figure}

\end{document}